\documentclass[twocolumn,english,showpacs,floatfix,amsmath,amssymb,prl,
superscriptaddress]{revtex4-1}
\usepackage[T1]{fontenc}
\usepackage[latin9]{inputenc}
\usepackage{babel}

\usepackage{graphicx}
\usepackage{epsfig,psfrag}
\usepackage{amssymb}
\usepackage{esint}
\usepackage{bm}
\usepackage{xcolor}



\begin{document}

\title{Iterative summation of path integrals for nonequilibrium molecular quantum transport}

\author{R. H\"utzen} 
\affiliation{\mbox{Institut f\"ur Theoretische Physik, Heinrich-Heine-Universit\"at, D-40225 D\"usseldorf, Germany}}

\author{S. Weiss} 
\affiliation{\mbox{Institut f\"ur Theoretische Physik, Heinrich-Heine-Universit\"at, D-40225 D\"usseldorf, Germany}}
\affiliation{I.\ Institut f\"ur Theoretische Physik, Universit\"at Hamburg, 
D-20355 Hamburg, Germany}

\author{M. Thorwart}
\affiliation{I.\ Institut f\"ur Theoretische Physik, Universit\"at Hamburg, 
D-20355 Hamburg, Germany}

\author{R. Egger} 
\affiliation{\mbox{Institut f\"ur Theoretische Physik, Heinrich-Heine-Universit\"at, D-40225 D\"usseldorf, Germany}}

\date{\today}
\begin{abstract}
We formulate and apply a nonperturbative numerical approach 
to the nonequilibrium current, $I(V)$, through a voltage-biased 
molecular conductor.  We focus on a single electronic 
level coupled to an unequilibrated vibration mode (Anderson-Holstein model),
which can be mapped to an effective three-state problem.  Performing an 
iterative summation of real-time path integral (ISPI) expressions, 
we accurately reproduce known analytical results in three different limits. 
We then study the crossover regime between those limits and show that the 
Franck-Condon blockade persists in the quantum-coherent low-temperature 
limit, with a nonequilibrium smearing of step features in the $IV$ curve.
\end{abstract}
\maketitle

{\it Introduction.---}Understanding quantum transport in 
nanoscale electronic systems with vibrational or mechanical  
(``phonon'') degrees of freedom is of topical interest in 
several areas of physics, including molecular 
electronics \cite{molel,pederson}, 
inelastic tunneling spectroscopy \cite{nitzan}, 
nanoelectromechanical systems \cite{nems},  
break junctions \cite{jan}, and suspended semiconductor or carbon-based 
nanostructures  \cite{Steele09,adrian2,ensslin,HuettelWitkamp}. 
The electron-phonon interaction allows to observe a rich variety of 
intriguing phenomena, such as negative differential conductance, the 
Franck-Condon blockade of transport, rectification, vibrational 
sidebands or steplike features in the current-voltage ($IV$) 
characteristics,  and current-induced heating or cooling.   
Already the simplest nonequilibrium ``Anderson-Holstein'' (AH) 
model, where the nanostructure corresponds to just one spinless electronic 
level coupled to a single oscillator mode, 
captures much of this richness \cite{Mitra,WangThoss2009}; for a review, see Ref.~\cite{nitzan}.  
Analytical approaches allow to understand the AH model in various 
corners of parameter space, but no controlled approximation,
let alone exact solution, connecting these corners seems in reach.  
One may expect that a unified picture is available from numerics.  
However, numerical renormalization
group \cite{cornaglia} or quantum Monte Carlo (QMC) calculations 
\cite{arrachea,RMP} are usually restricted to equilibrium.  
For the nonequilibrium AH model, Han \cite{han} employed an imaginary-time 
QMC approach followed by a double analytical continuation scheme;
unfortunately, the latter step is plagued by instabilities \cite{dirks}. 
A promising avenue for the AH model has recently been suggested by
real-time path-integral QMC simulations \cite{lothar,schiro}, where 
one directly computes the time-dependent current.
Such calculations have to deal with the infamous dynamical sign problem 
at long times, but in several parameter regions, especially when a 
secondary phonon bath is present, the stationary steady-state 
regime can be reached. 

In this work, we formulate and apply an alternative numerical 
approach, which in practice is useful unless both the temperature $T$ 
and the bias voltage $V$ are small.  It is also based on a
Keldysh path-integral formulation but does not involve stochastic 
sampling schemes and thus remains free from any sign problem.  
To that end, we extend the ``iterative summation of path integrals'' (ISPI) 
technique \cite{ispi} to the AH model.  
Technical aspects of the present approach, in particular 
our mapping to an effective three-state system via the ''spin-1 Hirsch-Fye 
transformation'' in Eq.~\eqref{aux} below, should also be of
interest to QMC schemes \cite{RMP}.  In essence, the ISPI method exploits that 
time correlations of the auxiliary three-state Keldysh variable, 
which arise after functional integration over the 
phonon and the (dot and lead) fermion degrees of freedom, 
can be truncated beyond a certain memory time $\tau_m$ 
when either $T$ or $V$ is finite.  Together with
a convergence scheme designed to eliminate systematic errors due to the
finiteness of $\tau_m$, such calculations allow to obtain numerically exact 
results.  The ISPI method has already been successfully applied to the
spinful Anderson model \cite{ispi,njp,segal}, where instead of the phonon 
a local charging interaction is present.  While we focus on the simplest
version of the AH model with a single unequilibrated \cite{foot1} 
phonon mode here, 
the conceptual generalization to include Coulomb interactions, more 
phonon modes, or several dot levels is straightforward. 
We benchmark our ISPI code against three different analytical approaches and 
then study the crossover between the respective regimes.

{\it AH Model.---}We consider the AH Hamiltonian, 
$H=H_m+H_t+H_l$, describing a molecular level with tunnel coupling ($H_t$)
to metallic source and drain contacts.  
Taking a single spinless dot level (fermion annihilation operator $d$) with 
energy $\epsilon$ and a boson mode (annihilation operator $b$) of 
frequency $\Omega$, the isolated molecule Hamiltonian is \cite{nitzan}
(we use units with $\hbar=k_B=1$)
\begin{equation}\label{hm}
H_m =  \Omega \ b^\dagger b + \left[ \epsilon+\lambda (b+b^\dagger) 
\right ] n_d 
\end{equation}
with $n_d=d^\dagger d$ and
the electron-phonon coupling strength $\lambda$.  The lead
Hamiltonian $H_l$ is taken in the standard wide-band approximation 
\cite{nazarov}, with relaxation processes assumed fast enough to 
have Fermi functions with temperature $T$  as electronic distributions;
their chemical potential difference defines the bias voltage $V$.  
The tunnel coupling then introduces the hybridization energy scales 
$\Gamma_L$ and $\Gamma_R$ for the left/right lead \cite{nazarov}.
For simplicity, we focus on the symmetric case in what follows, 
$\Gamma_L=\Gamma_R=\Gamma/2$. As observable of main interest, we study the 
steady-state current $I$ through the molecule.

{\it Analytical approaches.---}Before turning to a description of 
the ISPI scheme, let us briefly summarize the analytical approaches 
to the AH model that we employ to benchmark our method. 
(i)  For $\lambda/\Gamma\ll 1$,
perturbation theory in the electron-phonon coupling applies and 
yields a closed $IV$ expression for arbitrary values of 
all other parameters \cite{egger}.  We note that the solution
of the AH model with a very broad dot level \cite{dora,vinkler}
corresponds to this small-$\lambda$ regime.
(ii) For high temperatures, $T\gg \Gamma$, 
a description in terms of a rate equation is possible \cite{nazarov}.  
We here use the simplest sequential tunneling version with 
golden rule rates \cite{flensberg}.  For small $\lambda$, the 
corresponding results match those of perturbation theory, while in 
the opposite strong-coupling limit, the Franck-Condon blockade 
occurs and implies a drastic current suppression at 
low bias voltage \cite{ensslin,koch}. 
(iii) For small oscillator frequency, $\Omega\ll {\rm min}(\Gamma,eV)$,
the nonequilibrium Born-Oppenheimer (NEBO) approximation is controlled
and allows to obtain $I$ from a Langevin equation for the oscillator 
\cite{pistolesi,bode}.  For small $\lambda$, this approach is also
consistent with perturbative theory, while for high $T$, 
NEBO and rate equation results are found to agree.   

{\it  Keldysh path-summation formulation.---}We now start from the textbook
time-discretized coherent-state representation of the 
Keldysh generating functional \cite{kamenev,altland}.  The short-time
propagator on the forward/backward branch of the Keldysh contour, 
$e^{\mp i\delta_t H}$, where $\delta_t$ denotes the
discrete time step, then allows for a Trotter breakup, 
$e^{\mp i\delta_t H} =e^{\mp i\delta_t H_1} e^{\mp i\delta_t(H-H_1)}$,
with the systematic error in observables scaling $\sim\delta_t^2$.  
It is useful to choose $H_1= H_m-\Omega b^\dagger b$, 
see Eq.~\eqref{hm}, where the auxiliary relation
\begin{equation}\label{aux}
e^{\mp i \delta_t H_1}= 1 - n_d + n_d e^{-\lambda^2\delta_t^2/2}
e^{\mp i\delta_t\epsilon} e^{\mp i\delta_t\lambda b^\dagger} 
e^{\mp i\delta_t\lambda b} 
\end{equation}
allows to effectively decouple the electron-phonon interaction in terms of a 
three-state variable $s_{\eta}=0,\pm 1$ defined at each (discretized) time 
step $t_{j}$ along the forward/backward ($\sigma=\pm$) part of the 
Keldysh contour, where $\eta=(t_j,\sigma)$.
Below, we also use the notation $\eta\pm 1=(t_{j\pm 1},\sigma)$
with periodic boundary conditions on the Keldysh contour.
It is crucial for the construction of the coherent-state
functional integral that Eq.~\eqref{aux} is normal ordered.
The ``spin'' variable $s_\eta$ picks up the three terms in Eq.~\eqref{aux} 
and acts like a Hubbard-Stratonovich auxiliary field, similar to 
the Ising field employed in the Hirsch-Fye formulation of
the Anderson model \cite{ispi,fye}.  The bosonic (phonon) scalar field and the
fermionic (dot and lead electrons) Grassmann fields appearing in the 
Keldysh path integral are then effectively noninteracting but
couple to the time-dependent auxiliary spin variable.
Hence those fields can be integrated out analytically 
and the time-dependent current, $I(t_j)$, follows from a path-summation
formula for the generating functional, 
\begin{equation}\label{gen}
{\cal Z}=\sum_{\{ s_{\eta}=0,\pm 1 \}} {\rm det} D[\{s\}],
\end{equation}
where the matrix $D_{\eta\eta'}$ (in time and Keldysh space) depends on the 
complete spin path $\{s \}$. Specifically, we obtain $D=-iB(G_d^{-1}-\Sigma)$,
where $G_d^{-1}$ is the discretized inverse Green's function of 
the dot as in Refs.~\cite{ispi,kamenev} but with the modified 
spin-dependent matrix elements 
$\left[-iG^{-1}_d \right]_{\eta+1,\eta}=-s_{\eta}$.
The self-energy matrix $\Sigma$ describes the traced-out leads.
We find $\Sigma_{\eta\eta'}\ne 0$ only when $s_{\eta}=\pm 1$,
where it coincides with the usual (wide-band limit) expression \cite{nazarov}. 
Finally, the diagonal matrix $B$ (quoted here for $\epsilon=0$) with
\begin{equation}
B_{\eta\eta}=A_{s_{\eta}} e^{-\lambda^2\delta_t^2 \sum_{\eta'} \sigma \sigma'
[iG_{ph}]_{\eta,\eta'+1} | s_\eta s_{\eta'} |}
\end{equation}
encapsulates all phonon effects, where
$G_{ph}$ is the discretized phonon Green's function, 
see Ref.~\cite{kamenev}, and we used the notation
$A_{0}=1$ and $A_{\pm 1}=\pm (1/2) e^{-\lambda^2\delta_t^2/2}$.
By including a source term in $D$, it is straightforward to
numerically extract the time-dependent current $I(t_j)$ \cite{ispi}. 
For sufficiently long times $t_j$,  $I(t_j)$ reaches
a plateau yielding the steady-state current of interest.

{\it ISPI implementation.---}Starting from the formally exact path-summation 
formula [Eq.~\eqref{gen}], the ISPI algorithm can now be adapted from 
Ref.~\cite{ispi}:  For finite $T$ or $V$, matrix entries $D_{\eta\eta'}$ 
involving large time differences $|t-t'|$ are exponentially small.
We put these to zero beyond a memory time $\tau_m\equiv K\delta_t$, 
where $K$ denotes the number of time slices kept in the memory. 
The numerical computation of the memory-truncated path summation 
in Eq.~\eqref{gen} is then possible in an iterative way \cite{ispi} 
without additional approximations and, for given $\delta_t$ and $K$,
yields the steady-state current $I(\delta_t,K)$.  While this formulation is exact 
for $\delta_t\to 0$ and sufficiently large $K$ (long memory time), 
$K$ cannot be chosen arbitrarily large in practice and an extrapolation 
scheme is necessary \cite{ispi}.  Convergence of the extrapolation 
requires sufficiently high $T$ or $V$, for otherwise the necessary memory 
times are exceedingly long.  For the results below, 
we used $K\le 4$ and $0.3\le\Gamma\delta_t\le 0.35$.  The shown current
follows by averaging over the $\delta_t$-window, with error bars indicating 
the mean variance.  Additional ISPI runs for 
$0.18\le \Gamma\delta_t\le 0.22$ and $0.3\le \Gamma\delta_t\le 0.4$ 
were consistent with these results, and
we conclude that small error bars indicate that convergence has been reached.
For typical parameters and $K=4$, our ISPI code yielding 
$I(\delta_t,K)$ runs for $\approx 11$~CPU hours on 
a 2.93 GHz Xeon processor. 

\begin{figure}
\centering
\includegraphics[width=\columnwidth]{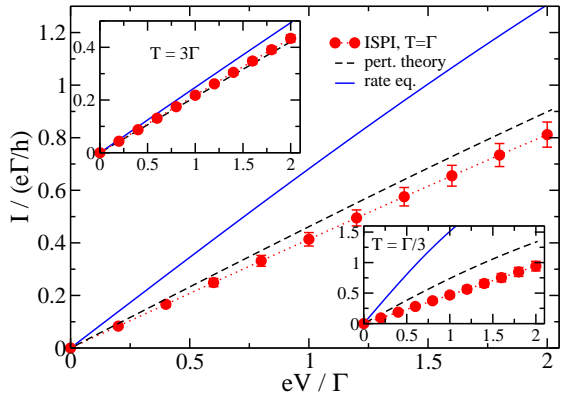}
\caption{(Color online) \label{fig1} 
Current $I$ (in units of $e\Gamma/h$)
vs bias voltage $V$ (in units of $\Gamma/e$) for
$\lambda=0.5\Gamma$, $\Omega=\Gamma$, $\epsilon=0$, and $T=\Gamma$. The
ISPI data are depicted as filled red circles, where the dotted red curve 
is a guide to the eye only and the error bars are explained in the main text.
We also show the results of perturbation theory in
$\lambda$ (dashed black curve) and of the rate equation 
(solid blue curve). The upper (lower) inset shows the 
corresponding result for $T=3\Gamma$ ($T=\Gamma/3$).}
\end{figure}

\begin{figure}
\centering
\includegraphics[width=\columnwidth]{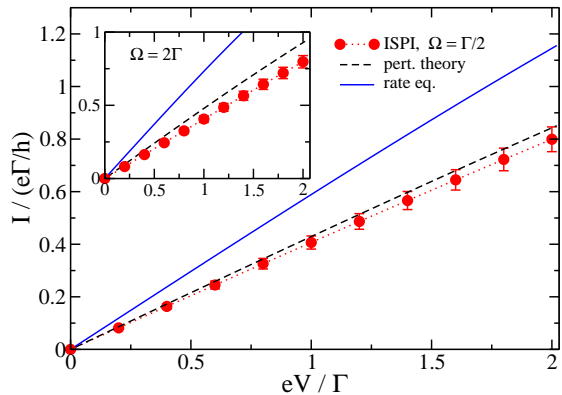}
\caption{(Color online) \label{fig2} 
Same as Fig.~\ref{fig1} but for $\Omega=0.5\Gamma$ (main panel) and $\Omega=
2\Gamma$ (inset), both for $T=\Gamma$.  }
\end{figure}

\begin{figure}
\centering
\includegraphics[width=\columnwidth]{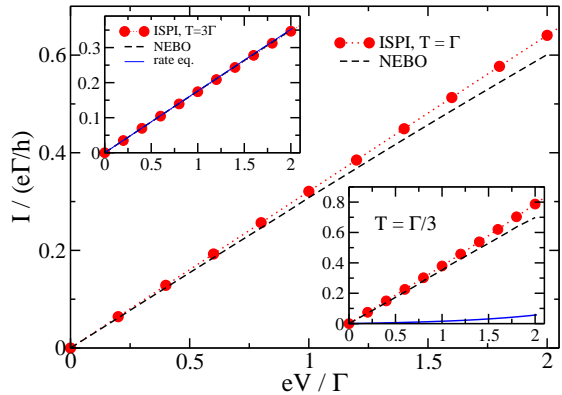}
\caption{(Color online) \label{fig3}
Same as Fig.~\ref{fig1} but for $\Omega=0.5\Gamma$ and
$\lambda=\Gamma$. The main panel is for $T=\Gamma$ and compares the ISPI 
results to NEBO predictions. The insets are for $T=3\Gamma$ and $T=\Gamma/3$, 
respectively, where also the rate equation results are shown. Notice that 
in contrast to ISPI, the rate equation predicts an 
unphysical current blockade for $T=\Gamma/3$. }
\end{figure}

{\it Benchmark checks.---}Next we show that the numerical ISPI results are
consistent with analytical theory for the $IV$ curves
in all three parameter limits mentioned above.  For clarity, we
focus on a resonant level with $\epsilon=0$ here.
Let us then start with the case of weak electron-phonon coupling,
$\lambda=0.5\Gamma$.  Figure \ref{fig1} compares our ISPI data for 
$\Omega=\Gamma$ to the respective results of perturbation theory 
in $\lambda$ and of the rate equation.  As expected, for this parameter choice, 
perturbation theory essentially reproduces the ISPI data.
The rate equation is quite accurate for high temperatures, 
but quantitative agreement with ISPI was obtained only 
for $T\agt 10\Gamma$. Note that the ISPI error bars increase when 
lowering $T$ due to the growing memory time ($\tau_m$) demands. 
The effect of changing the phonon frequency $\Omega$ 
is illustrated in Fig.~\ref{fig2}, taking $T=\Gamma$ but
otherwise identical parameters.  Again perturbation theory is well
reproduced.  Next, Fig.~\ref{fig3} shows ISPI results for a slow phonon mode, 
$\Omega=\Gamma/2$, with stronger electron-phonon coupling, $\lambda=\Gamma$. 
In that case, perturbation theory in $\lambda$ is not reliable and 
the rate equation is only accurate at the highest temperature ($T=3\Gamma$)
studied, cf.~the upper left inset of Fig.~\ref{fig3}.
However, we observe from Fig.~\ref{fig3} that for such a slow phonon mode, 
NEBO provides a good approximation for all temperatures and/or voltages of 
interest.  We conclude that the ISPI technique is capable of accurately 
describing three different analytically tractable parameter regimes.  

{\it Franck-Condon (FC) blockade.---}Next we address the limit of strong 
electron-phonon coupling $\lambda$, where the rate equation approach 
yields a FC blockade of the current for low bias and $T\gg \Gamma$ 
\cite{koch}. Sufficiently large $\lambda$ can be realized experimentally, and 
the FC blockade has indeed been observed in suspended carbon
nanotube quantum dots \cite{ensslin}.  For an unequilibrated phonon 
mode with intermediate-to-large $\lambda$, understanding the FC 
blockade in the quantum-coherent low-temperature regime, 
$T < \Gamma$, is an open theoretical problem. Here multiple phonon 
excitation and deexcitation effects imply a complicated (unknown)
nonequilibrium phonon distribution function, and the one-step tunneling 
interpretation in terms of FC matrix elements between 
shifted oscillator parabolas \cite{koch} is not applicable anymore. 
We here study this question using ISPI simulations, which 
automatically take into account quantum coherence effects.

\begin{figure}
\centering
\includegraphics[width=\columnwidth]{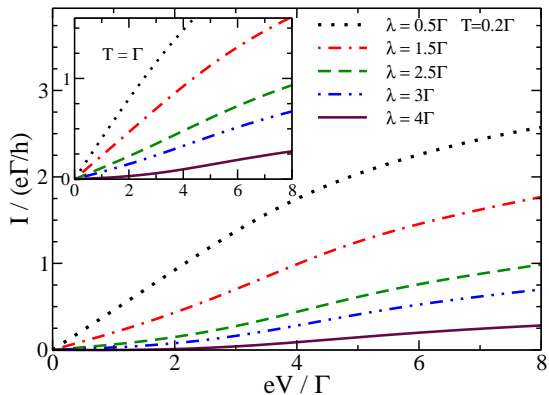}
\caption{(Color online) \label{fig4}
ISPI data for the $IV$ curves from weak ($\lambda=0.5\Gamma$) to strong 
($\lambda=4\Gamma$) electron-phonon coupling, with $\Omega=2\Gamma$. 
The main panel is for $T=0.2\Gamma$, the inset for $T=\Gamma$.
We used a dense voltage grid yielding smooth $IV$ curves.
Error bars are not shown but remain small, cp.~Fig.~\ref{fig1}.}
\end{figure}

In Fig.~\ref{fig4}, the crossover from weak to strong electron-phonon coupling 
$\lambda$ is considered. The inset shows $IV$ curves for $T=\Gamma$, where we 
observe a current blockade for low voltages once $\lambda\agt 2\Gamma$.
The blockade becomes more pronounced when increasing $\lambda$ 
and is lifted for voltages above the polaron energy
$\lambda^2/\Omega$ \cite{koch}. Remarkably, the FC blockade 
persists and becomes even sharper as one enters the quantum-coherent regime 
(here, $T=0.2\Gamma$), despite of the breakdown of the sequential tunneling 
picture. We also observe a nonequilibrium smearing of phonon step 
features in the $IV$ curves in Fig.~\ref{fig4},
 cf.~also Refs.~\cite{ensslin,koch}.

{\it Conclusions.---}We have extended the iterative simulation of path 
integrals (ISPI) technique to the Anderson-Holstein model, which is 
the simplest nonequilibrium model for quantum dots or molecules 
with an intrinsic bosonic (phonon) mode. Our formulation exploits a mapping 
to an effective three-state system and reproduces three analytical 
theories valid in different parameter regions. This extension of the 
ISPI approach then captures the full crossover between those limits 
unless both $T$ and $V$ are very small.  For strong 
electron-phonon coupling and an unequilibrated phonon mode, we find that
the Franck-Condon blockade becomes even more pronounced 
as one enters the quantum coherent regime.  

We thank A. Zazunov for helpful discussions. 
Financial support by the DFG (SPP 1243 and SFB 668) 
and by the ZIM (D\"usseldorf) is acknowledged.

\end{document}